\documentclass[11pt]{article}
\usepackage[margin=1in]{geometry}
\usepackage{amsmath,amssymb,amsthm}
\usepackage{mathtools}
\usepackage{booktabs}
\usepackage{setspace}
\usepackage{microtype}
\usepackage[hidelinks]{hyperref}
\usepackage{natbib}
\usepackage{graphicx}
\usepackage{xcolor}

\usepackage[compact]{titlesec}
\titlespacing*{\section}{0pt}{2.5\baselineskip}{1\baselineskip}
\usepackage[section]{placeins}

\newtheorem{proposition}{Proposition}
\newtheorem{definition}{Definition}
\newtheorem{assumption}{Assumption}
\theoremstyle{definition}

\title{\textbf{Algorithmic Intermediation and the International\\
Transmission of U.S. Monetary Policy}}
\author{Fernando Toledo \and Luis Dimotta Br\'e \and Gabriel Montes-Rojas}
\date{This draft: \today}

\begin{document}
\maketitle

\begin{abstract}
\noindent This paper examines how algorithmic and AI-driven fund management shapes the international transmission of U.S. monetary policy to emerging markets. It argues that the key source of instability is not algorithmic intermediation itself, but the similarity of models across funds. When algorithms rely on similar signals and make correlated errors, their trades reinforce one another and intensify capital-flow responses during periods of stress. When models are diverse, errors offset each other and algorithmic investors can stabilize flows. The paper develops a two-region macro-financial framework and tests its central prediction using equity portfolio flows to nineteen emerging markets from 2000 to 2024. The evidence shows that algorithmic herding amplifies outflows after U.S. monetary shocks only in high-volatility regimes, while faster adjustment alone has no comparable effect. The amplification is negative in sign across an extensive robustness battery and statistically significant under the baseline specification, with two qualifications we state openly: significance depends on defining the stress regime by the VIX rather than by calendar years, and it weakens when the 2008--09 crisis---the most informative stress episode---is excluded. The results imply that policy should focus on preserving model diversity rather than limiting the size of non-bank intermediation.


\end{abstract}

\newpage

\section{Introduction}

This paper asks whether algorithmic and AI-driven fund management changes the
international transmission of U.S. monetary shocks. The question matters for
international financial stability, and the answer is not obvious, because it
depends on how AI shapes financial decisions. On the one hand, common AI models
can raise market correlations and amplify stress episodes, and correlated
positioning already appears among the vulnerabilities that supervisors flag as a
source of systemic risk \citep{FSB2024ai}. On the other, adaptive strategies that
run on genuinely different architectures and data have been shown to supply
liquidity when conventional investors flee, stabilizing flows rather than
amplifying them. The same technology, then, looks like a source of fragility in
some hands and of resilience in others.

The answer lies not in the presence of algorithms but in
their \emph{similarity}. When many funds run correlated models---trained on
similar data, reading the same signals through the same pipelines---their
forecast errors move together, and their trades compound. When funds run diverse
models, their errors cancel, and the sector as a whole prices fundamentals more
accurately than any single participant. A single parameter, which we call
\textit{algorithmic homogeneity}, captures this distinction:
it is the fraction of forecast-error variance that algorithmic funds share.

Three facts about the current environment motivate both the modeling choice and
the empirical work. First, stress episodes in cross-border capital flows are
increasingly synchronized and fast: a small common trigger sets off a violent,
correlated response across otherwise unrelated markets, driven by joint
deleveraging rather than by country fundamentals. 
Second, the intermediaries capable of synchronized, model-driven trading are no longer
marginal. According to the Financial Stability Board, the non-bank financial
intermediation (NBFI) sector reached 49.1 percent of total global financial
assets in 2023, having grown 8.5 percent that year---more than double the pace
of the banking sector \citep{FSB2024gmr}. By 2024 the share had risen to 51.0
percent, or 256.8 trillion dollars \citep{FSB2025gmr}. The fastest-growing
component is the group that houses hedge funds, money market funds, and other
investment vehicles---precisely the entities most likely to run algorithmic
strategies. Third, the sensitivity of emerging-market capital flows to the global factor is not constant: fundamentals dominate in tranquil times, but in stress episodes the global factor becomes the dominant driver.

This paper develops a two-region DSGE model and derives
three results. First, homogeneity amplifies the transmission of U.S. monetary
shocks, but through the \emph{variance} of the aggregate lending response rather
than its mean: correlated errors survive aggregation while independent errors
wash out. Second, heterogeneity attenuates transmission, and in the limit of
fully diverse models the algorithmic sector becomes a perfectly informed,
stabilizing intermediary. Third, and most important for what follows, the
threshold level of homogeneity at which amplification sets in is
\emph{state-dependent}: it falls when local fundamentals deteriorate. This
delivers a sharp and testable prediction. The same algorithmic sector that
calms capital flows in normal times amplifies them in crises---not because
homogeneity itself changes, but because the threshold it must cross drops when
the economy is under stress.

We test this prediction on a panel of equity portfolio flows to nineteen
emerging markets over 2000--2024. Using a triple-interaction design that lets the
effect of U.S. monetary shocks vary with algorithmic behavior and with the
global stress regime, we find that the herding dimension of algorithmic
behavior amplifies capital outflows, and
that this amplification is confined to high-volatility periods, exactly as the
state-dependent threshold predicts. The negative sign of the amplification is
present in every specification of an extensive robustness battery, and it is
statistically significant under the baseline specification and under the
inference schemes best suited to the setting. We are equally explicit about the
limits of that battery: significance rests on two design choices we document
openly---defining the stress regime by the VIX rather than by calendar years,
and retaining the 2008--09 crisis, the single most informative stress episode. A
state-dependent mechanism that draws its identification from a small number of
stress quarters should be expected to lose precision precisely when those
quarters are redefined or removed, and we report those sensitivities rather than
obscure them.

The contribution is threefold, and it is worth stating precisely how it differs
from a fast-growing parallel literature on algorithmic homogeneity. Relative to
the literature on the Global Financial Cycle \citep{Rey2015, MARey2020}, which
documents that U.S. monetary policy drives a common factor in global asset
prices and flows, we provide a micro-founded mechanism---the aggregation of
correlated forecast errors---that explains why the strength of transmission
varies over time and across regimes. Relative to the literature on non-bank
financial intermediation, which has documented the sector's growth and its
procyclicality, we show that what matters for amplification is not the
institutional form of the intermediary but the behavioral correlation across
intermediaries. And relative to the recent structural work on algorithmic
monoculture \citep{MengChen2026, QiuHan2026}---which models the correlation of
machine behavior as a driver of aggregate systemic risk and validates it on
U.S. institutional holdings---our distinct object is the \emph{international
transmission} of monetary policy: we embed the correlation of forecast errors
in a two-region model of the U.S.--periphery cycle, an environment those
frameworks do not address. The sharpest point of departure is qualitative.
That literature derives an effect of algorithmic adoption on risk that is
monotone and increasing---more homogeneity, more fragility. Our central result
is instead a \emph{state-dependent sign reversal}: the same algorithmic sector
stabilizes capital flows when fundamentals are healthy and amplifies them only
once fundamentals deteriorate and the amplification threshold falls below
prevailing homogeneity. It is this conditional reversal, not the level effect of
homogeneity, that we take to the data.

The remainder of the paper is organized as follows.
Section~\ref{sec:lit} situates the contribution in the
literature. Section~\ref{sec:model} develops the two-region DSGE model, and
Section~\ref{sec:prop} derives the three propositions. Section~\ref{sec:robust}
establishes the robustness of the analytical results, and
Section~\ref{sec:calib} reports the calibration. Section~\ref{sec:empirics}
tests the model's central prediction on a panel of equity portfolio flows to
nineteen emerging markets. Section~\ref{sec:policy} draws
out the implications for macroprudential policy and concludes.

\section{Related Literature}\label{sec:lit}

Our contribution connects to three strands of literature.

The first is the literature on the Global Financial
Cycle. \citet{Rey2015} established that a common factor drives risky asset
prices, capital flows, and leverage worldwide, comoving with U.S. monetary
policy and the VIX, and \citet{MARey2020} showed through high-frequency
identification that U.S. monetary contractions tighten financial conditions
abroad even under floating exchange rates. This work motivates our
center--periphery structure and the state-dependence of Fact 3, but it treats
the intermediation sector as a black box. Our contribution is to open it: we
show that the \emph{composition} of intermediation---specifically the behavioral
correlation across algorithmic funds---determines whether a given U.S. shock is
amplified or dampened, providing a micro-founded reason why transmission
strength varies over time.

The second is the literature on heterogeneous financial intermediaries.
\citet{CoimbraRey2024} develop a macro-finance model in which the cross-section
of Value-at-Risk constraints generates a non-monotonic relationship between
interest rates and financial stability, with heterogeneity in \emph{risk-taking
capacity} as the driving force. Our model shares the emphasis on cross-sectional
heterogeneity but locates it elsewhere: not in how much risk intermediaries are
willing to bear, but in how correlated their signal-extraction errors are. This
distinction matters because our source of heterogeneity---model homogeneity
$\phi$---can be high even when risk appetites are identical, and it is $\phi$,
not risk capacity, that our propositions show to govern amplification. The
broader intermediary-based macro-finance tradition provides the natural point of
comparison for our state-dependent result. \citet{BrunnermeierSannikov2014} show
that an economy with a constrained financial sector exhibits highly non-linear
amplification and occasional volatile crises---their ``volatility paradox''---
while \citet{HeKrishnamurthy2013} model risk premia that rise non-linearly when
intermediary equity capital becomes scarce. In both, the non-linearity is driven
by the tightness of a balance-sheet or capital constraint. Our
Proposition~\ref{prop:threshold} delivers an analogous state-dependent
non-linearity from an entirely different primitive: not the scarcity of capital,
but the correlation of forecast errors across intermediaries. The two mechanisms
are complementary, and in a fuller model a binding capital constraint and rising
model homogeneity would reinforce one another.

The third is the literature on information choice and attention.
\citet{KVNV2016} provide a rational theory in which fund managers optimally
choose which signals to learn about, tying information choices to the business
cycle. We build on their information-theoretic foundation but invert one
assumption: in our setting funds observe the \emph{same} signals and differ in
the correlation of their noise. This is the formal device---anchored in the
exact law of large numbers for a continuum \citep{Sun2006}---through which
common forecast errors survive aggregation and produce herding. The
state-dependence of signal precision that drives our threshold result follows
the mechanism of \citet{OrlikVeldkamp2024}, in which real-time estimation of
non-normal distributions makes perceived uncertainty rise as outcomes move away
from the center.

Two further connections position the contribution. The concern that common AI
models raise market correlations and amplify stress is now explicit in official
assessments \citep{FSB2024ai}, but that literature is largely descriptive; we
supply the analytical mechanism behind the concern. And the amplification of
correlated flows in inelastic markets, quantified by \citet{GabaixKoijen2021},
is complementary to our channel: their price multiplier operates on the demand
side of asset markets, while ours operates on the correlation of intermediary
lending, and the two would compound in a fuller model. Finally, our mechanism is
distinct from the one in the emerging literature on strategic AI behavior.
\citet{DGJ2025} show that reinforcement-learning trading algorithms can learn to
collude---to sustain coordinated, supra-competitive behavior without explicit
communication. Their coordination is strategic and emerges from repeated
interaction; ours is statistical and emerges from shared training, requiring no
strategic intent. The two are complementary faces of the same underlying
concern, homogeneous machines behaving alike, and our contribution is to
provide the DSGE micro-foundation for the non-strategic, correlation-of-errors
channel that a general-equilibrium transmission model requires.

Closest to our object of study is a recent pair of structural papers on
algorithmic monoculture, and it is important to be explicit about where we
overlap and where we depart. \citet{MengChen2026} embed correlated algorithmic
signals in a Kyle-type market with performative feedback and endogenous
adoption, deriving a systemic-risk multiplier that grows superlinearly in the
AI-adoption share and validating it on the universe of SEC Form 13F holdings
with a shift-share instrument. \citet{QiuHan2026} construct a multi-agent model
whose AI traders share a representation layer, and argue that
\emph{representation} homogeneity---similarity in how agents encode market
states---compresses the space of forecast disagreement under stress even when
realized predictions look diverse in calm times. Our forecast-error homogeneity
$\phi$ is, in their taxonomy, the \emph{forecast-overlap} object rather than the
representation object; we take as our primitive precisely the quantity their
framework treats as secondary, but we do so in a setting neither addresses.
Three differences are substantive rather than terminological. First, both model
a single asset traded through a market maker; neither contains a central bank,
a monetary shock, or a center--periphery structure, so the international
transmission question we pose is outside their scope. Second, where they posit
signal correlation as an exogenous parameter, we \emph{derive} the survival of
common noise under aggregation from the exact law of large numbers for a
continuum \citep{Sun2006}, so that homogeneity operates through aggregation
alone and not through how funds weight their signals. Third, and most
important, their central prediction is monotone---adoption raises risk---whereas
Proposition~\ref{prop:threshold} delivers a reversal of sign across regimes.
This last difference also frames a limitation we return to in
Section~\ref{sec:empirics}: because their prediction is about tail risk, both
papers test a second moment directly, while our panel evidence identifies a
conditional-mean effect; aligning the two is the natural next step for the
empirical design rather than a settled matter.

\section{A Two-Region Economy with Two Types of Intermediary}\label{sec:model}

The model has a center---the United States---that sets monetary policy, and a
periphery---a representative emerging market---that receives cross-border
lending. The single novel ingredient is the intermediation sector: alongside
traditional funds that adjust lending sluggishly, a continuum of algorithmic
funds extracts noisy signals about U.S. policy and local fundamentals and lends
on the basis of those signals. A single parameter, the algorithmic homogeneity
$\phi \in [0,1]$, governs how correlated the funds' forecast errors are. The
paper's central result is that this one parameter determines whether algorithmic
intermediation amplifies or dampens the transmission of U.S. monetary shocks to
the periphery.

Throughout, we reserve $\phi$ for algorithmic homogeneity alone. The inverse
Frisch elasticity is denoted $\nu$, and the Taylor-rule coefficients are
$\varphi_\pi$ and $\varphi_y$, so that each symbol carries a single meaning.

\subsection{Households}

A representative household in the periphery chooses consumption $C_t$ and labor
$N_t$ to maximize
\begin{equation}
\mathbb{E}_0 \sum_{t=0}^{\infty} \beta^t
\left[ \frac{C_t^{1-\sigma}}{1-\sigma} - \chi \frac{N_t^{1+\nu}}{1+\nu} \right],
\label{eq:utility}
\end{equation}
where $\beta \in (0,1)$ is the discount factor, $\sigma > 0$ is the coefficient
of relative risk aversion (the inverse of the intertemporal elasticity of
substitution), $\nu > 0$ is the inverse Frisch elasticity of labor supply, and
$\chi > 0$ scales the disutility of work. Two features of \eqref{eq:utility}
matter for the transmission mechanism. First, $\sigma$ controls how strongly the
household smooths consumption across periods: a higher $\sigma$ makes the
household more reluctant to let consumption fluctuate, which flattens its
response to the interest-rate movements that a U.S. shock induces, and therefore
dampens the real-side amplification of any given financial shock. Second, $\nu$
governs how elastically labor responds to wage changes: a low $\nu$ (elastic
labor) lets the periphery absorb shocks through hours worked rather than through
consumption, again muting the propagation. These two curvature parameters are
the household-side channels through which the financial disturbance at the
center of the paper reaches the real economy. The household faces the budget
constraint
\begin{equation}
C_t + D_t + I_t = W_t N_t + R_{t-1} D_{t-1} + R^K_t K_{t-1} + \Pi_t
- \frac{\psi_I}{2}\left( \frac{I_t}{K_{t-1}} - \delta \right)^2 K_{t-1},
\label{eq:budget}
\end{equation}
and the capital law of motion $K_t = (1-\delta) K_{t-1} + I_t$. Here $D_t$ are
deposits paying the gross rate $R_t$, $I_t$ is investment, $W_t$ is the real
wage, $R^K_t$ is the rental rate of capital, $\Pi_t$ are profits rebated to the
household, $\delta \in (0,1)$ is the depreciation rate, and $\psi_I > 0$ governs
the quadratic cost of adjusting investment. This adjustment cost prevents
investment from jumping instantaneously to its
new optimum when a shock hits, and thereby generates the persistent, hump-shaped
investment response that the data display. Without it, the capital stock would
absorb shocks too quickly and the financial channel would leave no lasting
imprint on the real economy; $\psi_I$ is the parameter that ties the speed of
real adjustment to the financial disturbance. The optimality conditions
(derived step by step in the appendix) deliver the consumption Euler equation,
the labor-supply schedule, and the investment Euler equation:
\begin{align}
1 &= \beta\, \mathbb{E}_t \left[ \left( \frac{C_{t+1}}{C_t} \right)^{-\sigma}
R_t \right], \label{eq:euler_c}\\[0.3em]
\chi N_t^{\nu} C_t^{\sigma} &= W_t, \label{eq:labor}\\[0.3em]
Q_t &= \beta\, \mathbb{E}_t \left[ \left( \frac{C_{t+1}}{C_t} \right)^{-\sigma}
\left( R^K_{t+1} + Q_{t+1}(1-\delta) \right) \right], \label{eq:euler_i}
\end{align}
where $Q_t$ is Tobin's marginal $q$, the shadow value of installed capital.
Equation \eqref{eq:euler_c} is standard: the household equates the marginal
utility cost of saving one unit today to the discounted marginal utility gain of
consuming its gross return tomorrow---this is the channel through which the
policy rate $R_t$, moved by the U.S. shock, reshapes the periphery's saving and
consumption path. Equation \eqref{eq:labor} equates the marginal rate of
substitution between leisure and consumption to the real wage, closing the
labor-supply block. Equation \eqref{eq:euler_i} prices installed capital: the
household holds capital up to the point where its shadow value $Q_t$ equals the
discounted sum of next period's rental return and its resale value net of
depreciation. When $Q_t$ rises---because expected returns to capital improve or
the discount factor falls---investment expands, and it is through this valuation
equation that the credit conditions set by the intermediation sector feed into
real capital formation.

\subsection{Firms}

A representative firm produces output with the constant-returns technology
\begin{equation}
Y_t = A_t K_{t-1}^{\alpha} N_t^{1-\alpha},
\label{eq:production}
\end{equation}
where $\alpha \in (0,1)$ is the capital share and total factor productivity $A_t$
follows an AR(1) process, $\log A_t = \rho_A \log A_{t-1} + \varepsilon^A_t$. The
persistence $\rho_A$ determines how long a productivity disturbance lingers in
the periphery's fundamentals, and it therefore interacts directly with the
state-dependent signal precision of Section~\ref{sec:model}: the further and
longer $A_t$ (and with it the fundamental $F_t$) strays from steady state, the
noisier local signals become and the lower the amplification threshold falls.
Cost minimization yields the standard factor prices $W_t = (1-\alpha) Y_t / N_t$
and $R^K_t = \alpha Y_t / K_{t-1}$. Firms finance capital by borrowing from
intermediaries at the loan rate $\tilde{R}^K_t = R_t + \mu_t$, where the external
finance premium
\begin{equation}
\mu_t = \bar{\mu} \left( \frac{L_t}{\bar{L}} \right)^{\xi}, \qquad \xi > 0,
\label{eq:spread}
\end{equation}
rises with aggregate lending $L_t$ relative to its steady-state level $\bar{L}$.
Equation \eqref{eq:spread} is the pivotal link between the financial sector and
the real economy, and $\xi$ is arguably the single most consequential parameter
in the model. It measures how sharply the cost of external finance responds to
the volume of credit: when algorithmic funds withdraw lending in unison, $L_t$
falls, the premium $\mu_t$ rises, and firms face a higher loan rate
$\tilde{R}^K_t$ that chokes off investment. A large $\xi$ means the premium is
highly sensitive to credit volume, so a synchronized algorithmic retreat
produces a violent tightening of financial conditions; a small $\xi$ cushions
the same retreat. This is the transmission belt through which the correlation of
forecast errors, an object internal to the intermediation sector, becomes a
macroeconomic contraction in the periphery.

\subsection{Traditional funds}

Traditional funds are slow. They choose lending $L^T_t$ to maximize expected
profits net of a quadratic cost of adjusting the loan book:
\begin{equation}
\max_{L^T_t}\ \mathbb{E}_t \left[ \tilde{R}^K_t L^T_t - R_t L^T_t
- \frac{\psi_T}{2}\left( \frac{L^T_t}{L^T_{t-1}} - 1 \right)^2 L^T_{t-1} \right].
\label{eq:trad_problem}
\end{equation}
The first-order condition,
\begin{equation}
\mu_t = \psi_T \left( \frac{L^T_t}{L^T_{t-1}} - 1 \right),
\label{eq:trad_foc}
\end{equation}
makes traditional lending a smooth, backward-looking function of the premium:
the loan book expands only gradually, and the adjustment cost $\psi_T$ sets the
pace. This sluggishness is the foil against which algorithmic speed is defined.

\subsection{Algorithmic funds}

A continuum of algorithmic funds indexed by $j \in [0,1]$ observes noisy signals
about the U.S. monetary policy shock and about local fundamentals. Fund $j$'s
signals are
\begin{align}
s^{MP}_{j,t} &= MP_t + \nu^{MP}_{j,t}, \label{eq:sig_mp}\\
s^{LOC}_{j,t} &= F_t + \varepsilon^{LOC}_{j,t}, \label{eq:sig_loc}
\end{align}
where $MP_t$ is the policy shock, $F_t$ is the (economy-wide) periphery
fundamental, and the noise terms decompose into a common component and a
fund-specific idiosyncratic component:
\begin{equation}
\nu^{MP}_{j,t} = \sqrt{\phi}\, \zeta^{MP}_t + \sqrt{1-\phi}\, u^{MP}_{j,t},
\qquad
\varepsilon^{LOC}_{j,t} = \sqrt{\phi}\, \zeta^{LOC}_t + \sqrt{1-\phi}\, u^{LOC}_{j,t}.
\label{eq:noise}
\end{equation}
Here $\zeta^{MP}_t, \zeta^{LOC}_t$ are common shocks shared by all funds---the
signature of models trained on the same data with the same architecture---while
$u^{MP}_{j,t}, u^{LOC}_{j,t}$ are independent across funds.

\begin{assumption}[Noise normalization]
The common and idiosyncratic noise components have equal variance:
$\mathrm{Var}(\zeta^{k}_t) = \mathrm{Var}(u^{k}_{j,t}) = \sigma_k^2$ for
$k \in \{MP, LOC\}$.
\label{ass:norm}
\end{assumption}

Assumption \ref{ass:norm} carries analytical weight. Under \eqref{eq:noise}, the variance
of an individual fund's noise is
$\mathrm{Var}(\nu^{MP}_{j,t}) = \phi\,\sigma_{MP}^2 + (1-\phi)\,\sigma_{MP}^2$,
and the fraction of that variance attributable to the common component equals
$\phi$ \emph{only} when the two variances coincide. Without
Assumption \ref{ass:norm}, the fraction is
$\phi\sigma_\zeta^2 / [\phi\sigma_\zeta^2 + (1-\phi)\sigma_u^2] \neq \phi$, and
the interpretation of $\phi$ in Definition \ref{def:homog} fails. We therefore
impose it explicitly.

\begin{definition}[Algorithmic homogeneity]
The parameter $\phi \in [0,1]$ is the fraction of forecast-error variance that is
common across algorithmic funds. When $\phi = 0$, funds produce fully independent
errors (maximum diversity); when $\phi = 1$, all funds share the same error
(maximum homogeneity).
\label{def:homog}
\end{definition}

Each fund forms a minimum--mean-squared-error forecast of next-period periphery
output, weighting each signal by its Bayesian precision:
\begin{equation}
\mathbb{E}^A_{j,t}[Y_{t+1}] = \theta^{MP}\, a\, s^{MP}_{j,t}
+ \theta^{LOC}\, b\, s^{LOC}_{j,t},
\label{eq:forecast}
\end{equation}
where $a, b$ are the loadings of output on the policy and fundamental states and
the weights $\theta^{MP}, \theta^{LOC}$ are the signal-to-total precision ratios.
Under Assumption \ref{ass:norm} the variance of each fund's signal noise is
$\sigma_k^2$, \emph{independent of $\phi$}. It follows that the Bayesian weights
themselves do not depend on $\phi$:
\begin{equation}
\frac{\partial \theta^{MP}}{\partial \phi}
= \frac{\partial \theta^{LOC}}{\partial \phi} = 0.
\label{eq:theta_indep}
\end{equation}
This is a substantive result, verified symbolically, and it clarifies the
mechanism: homogeneity does not operate by changing how funds weight their
signals. It operates entirely through aggregation, as we now show.

The precision of the local signal is state-dependent. Following the mechanism in
\citet{OrlikVeldkamp2024}---whereby real-time estimation of a distribution with
non-normal tails makes perceived uncertainty rise as outcomes move away from the
center---we let
\begin{equation}
\sigma_{LOC,t}^2 = \sigma_0^2 + \kappa_F \left( F_t - \bar{F} \right)^2,
\qquad \kappa_F > 0.
\label{eq:state_prec}
\end{equation}
Local signals become noisier when fundamentals are far from their steady state,
in either direction---deep recession or overheating. This single feature drives
the regime dependence of Proposition \ref{prop:threshold}.

Each fund chooses lending $L^A_{j,t}$ under a static mean--variance objective,
yielding the linear rule
\begin{equation}
L^A_{j,t} = \bar{L}_A - \gamma_1 \hat{\mu}_t
+ \gamma_2\, \mathbb{E}^A_{j,t}[Y_{t+1}],
\label{eq:algo_rule}
\end{equation}
where $\hat{\mu}_t$ is the log-deviation of the premium, $\gamma_1 > 0$ is the
sensitivity of lending to the premium, and $\gamma_2 > 0$ is the sensitivity to
the output forecast. The choice of a static objective for algorithmic funds---in
contrast to the intertemporal problem \eqref{eq:trad_problem} of traditional
funds---is deliberate and encodes the paper's central contrast: algorithmic
funds react to current signals without internalizing adjustment costs, which is
what makes them fast.

Aggregating \eqref{eq:algo_rule} across $j$ and applying the exact law of large
numbers for a continuum \citep{Sun2006}, the idiosyncratic terms
$u^{MP}_{j,t}, u^{LOC}_{j,t}$ integrate to zero, while the common terms survive:
\begin{equation}
L^A_t = \bar{L}_A - \gamma_1 \hat{\mu}_t + \gamma_2 \left[
\theta^{MP} a \left( MP_t + \sqrt{\phi}\, \zeta^{MP}_t \right)
+ \theta^{LOC} b \left( F_t + \sqrt{\phi}\, \zeta^{LOC}_t \right) \right].
\label{eq:algo_agg}
\end{equation}
Equation \eqref{eq:algo_agg} is the heart of the model. The common noise
$\sqrt{\phi}\,\zeta$ does not vanish under aggregation: it moves every fund in
the same direction at once. This is the formal content of herding. When
$\phi = 0$, the common terms disappear and the aggregate algorithmic forecast is
exact---the funds behave as a single, perfectly informed, stabilizing
intermediary.

\subsection{Market clearing and monetary policy}

Total lending is $L_t = L^T_t + L^A_t$, with households allocating a
steady-state share $\omega$ of deposits to traditional funds and $1-\omega$ to
algorithmic funds. (We treat $\omega$ as a fixed structural share; endogenizing
the portfolio choice is a straightforward extension that does not alter the
transmission results.) The U.S. central bank sets its policy rate according to an
inertial Taylor rule:
\begin{equation}
R_t = \rho_R R_{t-1} + (1-\rho_R)
\left[ \bar{R} + \varphi_\pi \pi_t + \varphi_y y_t \right]
+ \varepsilon^{MP}_t.
\label{eq:taylor}
\end{equation}
The rule combines two ideas. The bracketed term is the standard \citet{Taylor1993}
prescription: the central bank raises the nominal rate above its long-run level
$\bar{R}$ whenever inflation $\pi_t$ or the output gap $y_t$ rises above target,
with $\varphi_\pi$ and $\varphi_y$ measuring the strength of each response. The
requirement $\varphi_\pi > 1$---the Taylor principle---ensures that the real rate
rises when inflation does, which \citet{Woodford2003} and \citet{Gali2015} show
is necessary for a determinate, stable equilibrium; we set $\varphi_\pi = 2.0$,
consistent with the estimates of \citet{CGG2000} for the post-Volcker period.
The remaining term, $\rho_R R_{t-1}$, introduces interest-rate smoothing: the
central bank moves the rate only partially toward its target level each period,
adjusting gradually rather than in discrete jumps. This inertia is a robust
empirical feature of policy and has an optimizing rationale---\citet{Woodford2003}
shows that history-dependent, inertial rules arise naturally when the central
bank internalizes the effect of expected future rates on current conditions. The
smoothing parameter $\rho_R \in [0,1)$ governs the speed of adjustment; a higher
$\rho_R$ means a more gradual policy path. The term $\varepsilon^{MP}_t$ is the
monetary policy shock whose international transmission the paper studies: it is
the innovation in the policy rate that is orthogonal to the systematic response
to inflation and the output gap, and it is the disturbance that algorithmic funds
in the periphery must read through their noisy signals.

\section{Analytical Propositions}\label{sec:prop}

We now state the three propositions. Each is followed by a discussion. Proofs
are collected in the appendix. The three results map directly onto the three
stylized facts. Proposition~\ref{prop:amp} formalizes \emph{Synchronization}: it
identifies the survival of common forecast errors under aggregation as the
mechanism that turns homogeneous machines into a source of amplified,
correlated flows. Proposition~\ref{prop:att} shows the converse---that diversity
across models is stabilizing---which is why \emph{Scale} alone does not imply
fragility: a large algorithmic sector is dangerous only when it is homogeneous.
And Proposition~\ref{prop:threshold} reproduces \emph{State dependence}: the same
sector attenuates transmission in tranquil times and amplifies it in crises,
because the threshold separating the two regimes falls as fundamentals
deteriorate.

\begin{proposition}[Amplification under homogeneity]
Consider the response of aggregate algorithmic lending to a one-standard-
deviation contractionary monetary shock. The \emph{mean} response is independent
of $\phi$, but the \emph{variance} of the aggregate response is proportional to
$\phi$:
\begin{equation}
\mathrm{Var}\!\left( \frac{\partial L^A_t}{\partial MP_t} \right)
= \left( \gamma_2 \theta^{MP} a \right)^2 \phi\, \sigma_{MP}^2.
\end{equation}
The amplification factor---the ratio of the standard deviation of the aggregate
response under high homogeneity $\phi_H$ to that under low homogeneity
$\phi_L$---is
\begin{equation}
\mathcal{A}(\phi_H, \phi_L) = \sqrt{\phi_H / \phi_L}.
\label{eq:ampfactor}
\end{equation}
\label{prop:amp}
\end{proposition}

\begin{figure}[t]
\centering
\includegraphics[width=0.82\textwidth]{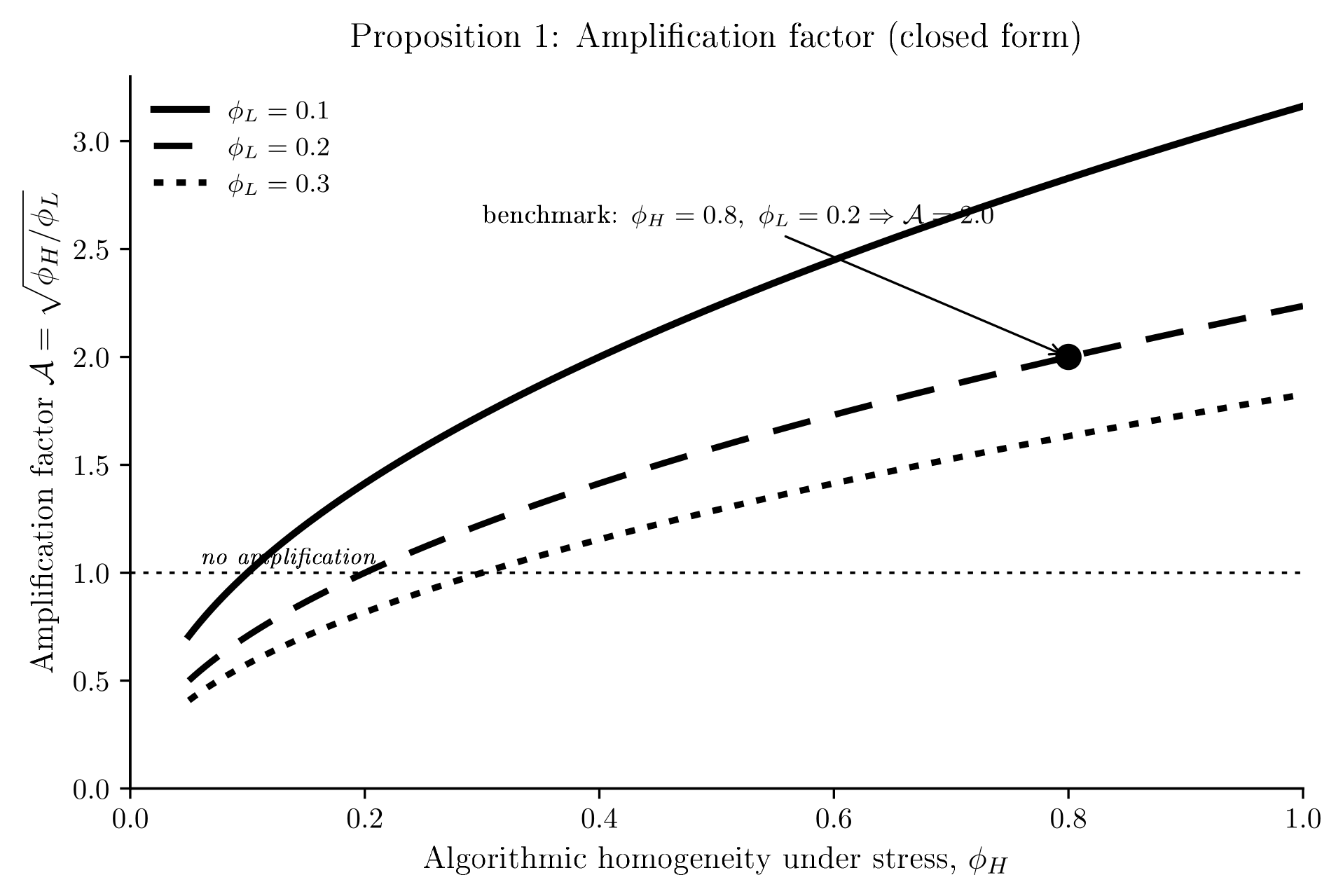}
\caption{The amplification factor $\mathcal{A} = \sqrt{\phi_H/\phi_L}$ of
Proposition \ref{prop:amp}, for three values of the normal-times homogeneity
$\phi_L$. The benchmark comparison ($\phi_H = 0.8$, $\phi_L = 0.2$) yields
$\mathcal{A} = 2.0$.}
\label{fig:prop1}
\end{figure}

A monetary shock has a deterministic effect on the
average fund's forecast, and that effect does not depend on how correlated the
funds' errors are. What homogeneity changes is the \emph{dispersion} of the
aggregate response. When errors are independent they wash out and the aggregate
is smooth; when they are correlated they compound, and the aggregate lending
response inherits a common component that scales with $\sqrt{\phi}$. Amplification
is therefore a statement about the volatility of capital flows, not about their
average level---which is precisely why it shows up in stress episodes rather than
in means. Evaluated at $\phi_H = 0.8$ and $\phi_L = 0.2$, the factor is exactly
$2.0$; across the plausible range it runs from about $1.4$ to $3.0$
(Table \ref{tab:sens}). Figure \ref{fig:prop1} plots the closed-form factor: it
is a transparent function of the two homogeneity levels, rising with the stress
homogeneity $\phi_H$ and falling with the normal-times homogeneity $\phi_L$, and
crossing unity---no amplification---only when the two coincide.

\begin{proposition}[Attenuation under heterogeneity]
The variance of the aggregate algorithmic forecast error is proportional to
$\phi$. As $\phi \to 0$, the aggregate forecast becomes unbiased with zero
variance, and the herding channel vanishes:
\begin{equation}
\lim_{\phi \to 0} \mathrm{Var}\!\left( \bar{\varepsilon}^A_t \right) = 0.
\end{equation}
\label{prop:att}
\end{proposition}

 With diverse models, the mistakes of one fund are
offset by the opposite mistakes of another. In the aggregate, the errors cancel
and the algorithmic sector delivers an exact reading of fundamentals. Diverse
algorithmic intermediation is thus not merely harmless---it is actively
stabilizing, supplying countercyclical liquidity because it prices fundamentals
correctly when correlated investors do not.

\begin{proposition}[State-dependent amplification threshold]
Let the net effect of algorithmic intermediation on the comovement between
capital flows and monetary shocks be the difference between an
information-processing (attenuating) channel that scales with local signal
precision $\tau_L = 1/\sigma_{LOC,t}^2$ and a herding (amplifying) channel that
scales with $\phi$. There is a threshold
\begin{equation}
\phi^* = \frac{\mathcal{A}_{\text{info}}\, \tau_L}{\mathcal{A}_{\text{herd}}}
= \frac{\mathcal{A}_{\text{info}}}
{\mathcal{A}_{\text{herd}}\left( \sigma_0^2 + \kappa_F\, \Delta F^2 \right)},
\qquad \Delta F \equiv |F_t - \bar{F}|,
\end{equation}
such that the net effect is attenuating for $\phi < \phi^*$ and amplifying for
$\phi > \phi^*$. The threshold falls as fundamentals deteriorate:
\begin{equation}
\frac{\partial \phi^*}{\partial \Delta F}
= - \frac{2\, \mathcal{A}_{\text{info}}\, \kappa_F\, \Delta F}
{\mathcal{A}_{\text{herd}}\left( \sigma_0^2 + \kappa_F \Delta F^2 \right)^2}
< 0.
\end{equation}
\label{prop:threshold}
\end{proposition}

\begin{figure}[t]
\centering
\includegraphics[width=0.82\textwidth]{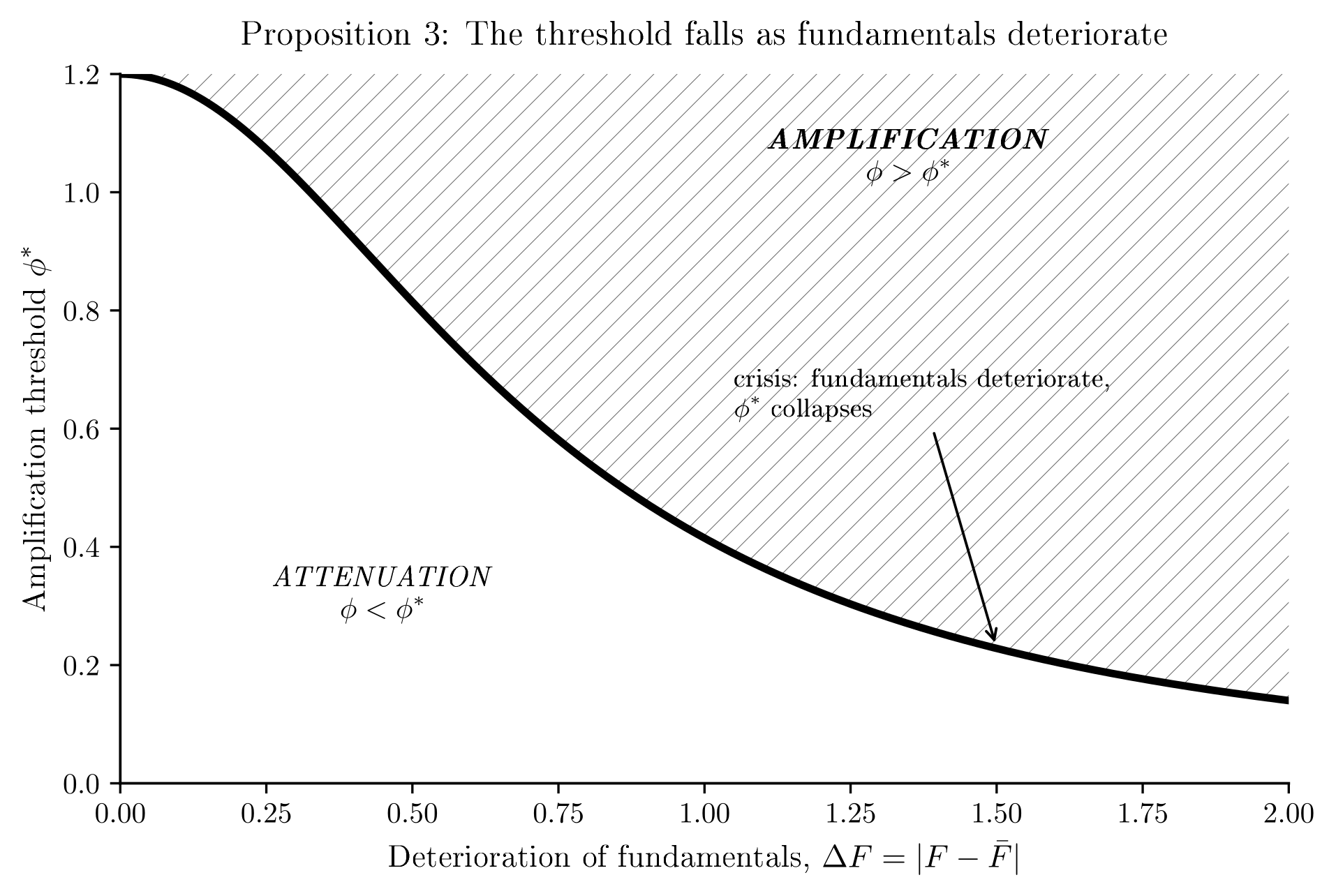}
\caption{The state-dependent amplification threshold $\phi^*$ of Proposition
\ref{prop:threshold}. As fundamentals deteriorate ($\Delta F$ rises), $\phi^*$
falls, and a fixed homogeneity level crosses from the attenuation region into
the amplification region.}
\label{fig:prop3}
\end{figure}

The threshold in the previous proposition is the level of homogeneity at which
the two channels exactly offset. In normal times, local signals are precise, the
information channel is strong, and $\phi^*$ is high---so realistic levels of
homogeneity leave transmission in the attenuating region. In a crisis,
fundamentals move far from steady state, local signals lose precision, the
information channel weakens, and $\phi^*$ collapses. Even moderate homogeneity
now exceeds the threshold, and the same algorithmic sector that stabilized flows
in calm times amplifies them under stress. This sign reversal---attenuation in
normal periods, amplification in crises---is the model's sharpest prediction and
the one the empirical section is designed to test. Note that the reversal is a
property of the threshold crossing, not of any change in $\phi$ itself:
homogeneity can be constant while its consequences flip with the state of the
economy. Figure \ref{fig:prop3} makes the mechanism visible: the threshold
$\phi^*$ traces a falling curve in the deterioration of fundamentals, dividing
the $(\Delta F, \phi)$ plane into an attenuation region below and an
amplification region above. A fixed level of homogeneity that sits in the
attenuation region in calm times is overtaken by the collapsing threshold as
fundamentals worsen.

\section{Robustness of the Analytical Results}\label{sec:robust}

Three assumptions underlie the propositions: the equal-variance normalization of
Assumption~\ref{ass:norm}, the reduced-form spread \eqref{eq:spread}, and the
quadratic form of the state-dependent precision \eqref{eq:state_prec}. We now
show that none of the three central results---amplification, the amplification
factor, and the falling threshold---depends on the specific form of these
assumptions.

Consider first relaxing the equal-variance normalization. Assumption~\ref{ass:norm}
sets $\sigma_\zeta^2 = \sigma_u^2$, which makes $\phi$ the exact fraction of
common forecast-error variance. Suppose instead that the two variances differ,
and let $\rho \equiv \sigma_\zeta^2 / \sigma_u^2$ be their ratio. The variance of
the aggregate common component becomes proportional to $\phi\rho$ rather than to
$\phi$, but the amplification factor---the ratio of the standard deviation of
the aggregate response under high versus low homogeneity---is
\begin{equation}
\mathcal{A} = \sqrt{\frac{\phi_H\, \rho}{\phi_L\, \rho}}
= \sqrt{\frac{\phi_H}{\phi_L}},
\end{equation}
in which $\rho$ cancels exactly. The amplification result of
Proposition~\ref{prop:amp} is therefore invariant to the equal-variance
normalization. What the normalization buys is interpretive convenience, not the
result itself: without it, $\phi$ ceases to be literally the fraction of common
variance and becomes a monotone proxy for homogeneity, while every comparative
static goes through unchanged.

The real-side transmission
runs through the external-finance premium \eqref{eq:spread}, whose elasticity to
credit volume is exactly $\xi$: differentiating, $\mathrm{d}\log\mu_t /
\mathrm{d}\log L_t = \xi$. A synchronized algorithmic withdrawal of a given size
therefore produces a premium increase that is strictly increasing in $\xi$, so
the real amplification inherits the sign and monotonicity of the financial
amplification for every admissible $\xi > 0$. The mechanism does not rely on a
knife-edge value of the elasticity; it operates across its entire range,
strengthening as the credit market becomes more inelastic---precisely the regime
that the inelastic-markets literature identifies as empirically relevant.

Proposition~\ref{prop:threshold}
was derived under a quadratic precision loss, $\sigma_{LOC,t}^2 = \sigma_0^2 +
\kappa_F \Delta F^2$. Replace the quadratic with a general power,
$\sigma_{LOC,t}^2 = \sigma_0^2 + \kappa_F \Delta F^{\,p}$ for any $p > 0$. The
threshold becomes $\phi^* = \mathcal{A}_{\text{info}} / [\mathcal{A}_{\text{herd}}
(\sigma_0^2 + \kappa_F \Delta F^{\,p})]$, and its derivative,
\begin{equation}
\frac{\partial \phi^*}{\partial \Delta F}
= - \frac{\mathcal{A}_{\text{info}}\, \kappa_F\, p\, \Delta F^{\,p-1}}
{\mathcal{A}_{\text{herd}}\left( \sigma_0^2 + \kappa_F \Delta F^{\,p} \right)^2}
< 0,
\end{equation}
is negative for every $p > 0$ and every $\Delta F > 0$. The sign reversal that
Proposition~\ref{prop:threshold} predicts---attenuation in normal times,
amplification in crises---is thus a property of the mechanism, not of the
quadratic specification: any precision loss that is increasing in the distance
of fundamentals from steady state delivers the same falling threshold. This is
the sense in which the paper's sharpest prediction is structural rather than
functional-form dependent.

\section{Calibration}\label{sec:calib}

Table \ref{tab:calib} reports the quarterly calibration. We distinguish sources
carefully. Values labeled ``standard'' are conventional in the New Keynesian
literature; values labeled ``calibrated'' are set to match a stated target and
are not drawn from a specific external estimate; and values with a citation are
taken from the cited source. Two parameter choices warrant explanation. First,
the Taylor-rule inflation coefficient is set following
\citet{CGG2000}, who estimate a forward-looking reaction function in which the
response to expected inflation in the Volcker--Greenspan period is well above
unity; we therefore use $\varphi_\pi = 2.0$ rather than the Taylor (1993) value
of $1.5$. Second, the state-dependence parameter $\kappa_F$ is calibrated to the
dispersion of emerging-market forecast errors; the mechanism is motivated by
\citet{OrlikVeldkamp2024}, but the numerical value is not taken from that paper.

\begin{table}[ht]
\centering
\caption{Baseline calibration (quarterly)}
\label{tab:calib}
\begin{tabular}{llll}
\toprule
Parameter & Value & Description & Source \\
\midrule
$\beta$ & 0.99 & Discount factor & Standard \\
$\sigma$ & 2.0 & Risk aversion & \citet{Gali2015} \\
$\nu$ & 1.0 & Inverse Frisch elasticity & \citet{Gali2015} \\
$\delta$ & 0.025 & Depreciation rate & Standard \\
$\alpha$ & 0.33 & Capital share & U.S. national accounts \\
$\psi_T$ & 0.80 & Traditional lending adjustment cost & Calibrated \\
$\gamma_1$ & 0.16 & Algorithmic sensitivity to premium & Calibrated \\
$\gamma_2$ & 0.80 & Algorithmic sensitivity to forecast & Calibrated \\
$\sigma_{MP}^2$ & 0.01 & MP signal noise variance & HF surprise data \\
$\sigma_0^2$ & 0.05 & Baseline local noise & EM forecast errors \\
$\kappa_F$ & 0.10 & State-dependent noise curvature & Calibrated$^{a}$ \\
$\phi$ & $[0,1]$ & Algorithmic homogeneity & Varies \\
$\omega$ & 0.70 & Traditional-fund share & FSB NBFI data$^{b}$ \\
$\xi$ & 0.20 & Spread elasticity & Credit-spread data \\
$\rho_R$ & 0.75 & Interest-rate smoothing & \citet{CGG2000} \\
$\varphi_\pi$ & 2.0 & Taylor rule: inflation & \citet{CGG2000} \\
$\varphi_y$ & 0.50 & Taylor rule: output gap & \citet{CGG2000} \\
\bottomrule
\end{tabular}

\vspace{0.2em}
\centerline{\begin{minipage}{0.85\textwidth}
{\footnotesize \color{black}
\textit{Notes.} $^{a}$ Mechanism motivated by \citet{OrlikVeldkamp2024}; the
value is calibrated to the dispersion of emerging-market forecast errors.
$^{b}$ The traditional-fund share $\omega$ is set to the complement of the NBFI
share of credit intermediation reported in the FSB Global Monitoring Report on
Non-Bank Financial Intermediation.}
\end{minipage}}
\end{table}

The amplification factor of Proposition \ref{prop:amp} depends only on the two
homogeneity levels being compared, not on the rest of the calibration. Table
\ref{tab:sens} reports it across a grid. It exceeds three only at implausibly
extreme homogeneity gaps ($\phi_H = 0.9$, $\phi_L \approx 0.1$); at the benchmark
comparison ($\phi_H = 0.8$, $\phi_L = 0.2$) it equals $2.0$. We report the factor
as an analytical object rather than committing to a single number, since it is a
transparent function of the homogeneity levels.

\begin{table}[ht]
\centering
\caption{Amplification factor $\mathcal{A} = \sqrt{\phi_H/\phi_L}$}
\label{tab:sens}
\begin{tabular}{lccc}
\toprule
& $\phi_L = 0.1$ & $\phi_L = 0.2$ & $\phi_L = 0.3$ \\
\midrule
$\phi_H = 0.6$ & 2.45 & 1.73 & 1.41 \\
$\phi_H = 0.7$ & 2.65 & 1.87 & 1.53 \\
$\phi_H = 0.8$ & 2.83 & 2.00 & 1.63 \\
$\phi_H = 0.9$ & 3.00 & 2.12 & 1.73 \\
\bottomrule
\end{tabular}
\end{table}

\section{Empirical Evidence}\label{sec:empirics}

The model leaves us with one prediction that can be taken to data. Algorithmic
homogeneity should amplify the transmission of U.S. monetary shocks to
emerging-market capital flows, and it should do so only in stress regimes, once
the amplification threshold $\phi^*$ has fallen below prevailing homogeneity
(Proposition~\ref{prop:threshold}). We test that prediction on a panel of equity
portfolio flows. The herding dimension of algorithmic behavior, which is our
empirical counterpart of $\phi$, does amplify capital outflows after U.S.
monetary shocks, and the amplification appears only when global financial stress
is high. The sign of the effect is robust. Its statistical significance is not,
and it turns on two design choices that we would rather state here than have a
reader find on their own.

\subsection{Data, variables, and sample}

The panel covers 19 emerging-market economies at quarterly frequency from
2000Q1 to 2024Q4. The dependent variable, $Y_{it}$, is the equity portfolio flow
to country $i$ in quarter $t$, drawn from the IMF Balance of Payments statistics
(BPM6). The estimation sample contains $N = 1{,}636$ country--quarter
observations.

The variable $Y_{it}$ records how much equity investment foreign investors
channelled into, or pulled out of, country $i$ over the quarter. 
$Y_{it}$ is built in two steps that replace the raw panel's annual-repeated
series with genuine quarterly variation. Portfolio equity flows are downloaded
from the IMF Balance of Payments statistics (BPM6) at quarterly frequency, on
the liabilities side of the portfolio-investment-equity account (indicator
\texttt{P\_F5}, accounting entry \texttt{L\_NIL\_T}, the net incurrence of
liabilities); \texttt{P\_F5} (equity plus investment-fund shares) is preferred
to the narrower \texttt{P\_F51} because it maximizes historical coverage back
to 2000Q1, and the two coincide almost exactly for economies where fund shares
are a small part of the account. A positive value means nonresidents are
increasing their equity claims on country $i$---a net portfolio inflow. The
resulting series, in current USD, is divided by $10^9$ so that $Y_{it}$ is
measured in USD billions. Two further points  on how the coefficients read. The first is that
$Y_{it}$ is a flow and not a stock, so what we estimate is how a monetary shock
moves the pace at which capital enters or leaves, not the level of holdings that
have already accumulated. The second is that the BPM6 framework fixes both the
sign convention and the boundary of what counts as a portfolio flow.
\citet{KoepkePaetzold2024} document these conventions and argue that
balance-of-payments portfolio-flow data are the right measure for the kind of
macroeconomic and external-financing question we are asking, in contrast to the
high-frequency fund-flow proxies such as EPFR, which cover only the mutual-fund
and ETF subset of investors. The BPM6 series therefore lines up with the object
the model actually speaks to, namely the total cross-border equity flow the
sector intermediates, rather than a partial and selection-prone slice of it.

Algorithmic behavior is captured by \emph{A\_herding}, which is built from the BIS Locational Banking Statistics in three steps. First, for each country and quarter we compute the percentage change in
total (bank and non-bank) external asset positions relative to the previous
quarter. Second, we average that same percentage change across every country in
the panel in a given quarter to obtain a global benchmark, the common
component of position growth that quarter. Third, \emph{A\_herding}$_{it}$ is
the rolling Pearson correlation, over the trailing eight quarters (the current
quarter and the seven before it, requiring at least six non-missing paired
observations), between the country's own percentage change and the global
benchmark. A country whose position changes closely follow the global benchmark
over that two-year window has a high \emph{A\_herding}, meaning its intermediaries are, in the aggregate, behaving like everyone else; a country whose changes are not correlated with the benchmark has a low or negative value. The proxy is lagged for one quarter, $A\_herding_{i,t-1}$, to avoid using information
contemporaneously with $r_t$. \emph{A\_herding} is the empirical object the
model identifies as the source of amplification.

The monetary shock $r_t$ is the high-frequency U.S. monetary policy surprise from
the San Francisco Fed's USMPD series, sign-flipped so that a positive value is an
expansionary surprise. 
 We want to be clear that this is not a proxy of
convenience. It measures the same kind of statistical object the theory puts at
the center of the mechanism. Recall that in the aggregation step of
Section~\ref{sec:model} the idiosyncratic components of algorithmic forecast
errors wash out under the exact law of large numbers, while the common component
survives and scales the variance of aggregate lending by $\phi$. Homogeneity, in
other words, works through the cross-sectional correlation of what funds do, not
through the level or the speed of any single fund's activity. The finance
literature measures herding in the same way, as the tendency of investors to move
into and out of the same positions at the same time \citep{Sias2004}. That
literature also draws the distinction we need, between spurious herding, where
funds simply react alike to common fundamentals, and intentional herding, where
the clustering comes from imitation itself; the distinction is set out in
\citet{BikhchandaniSharma2000} and has its theoretical roots in
\citet{Banerjee1992} and \citet{ScharfsteinStein1990}. Because our specification
interacts the proxy with an identified monetary shock and conditions on the
stress regime and on country fundamentals, it filters out the mechanical
comovement that common exposure alone would produce, and what remains is the
correlated-behavior component that stands in for $\phi$. The eight-quarter window
is a compromise. It is long enough to estimate a correlation with usable
precision and short enough to let the measure move with the state of the world,
which Proposition~\ref{prop:threshold} needs it to do, and the one-quarter lag
keeps the synchrony we measure ahead of the flow response it is meant to explain.
What the theory identifies as the source of amplification is this correlation,
not the raw velocity of adjustment, which is why we build the proxy to capture
synchrony rather than speed. 

The stress regime is a dummy, $\mathrm{Crisis}_t = \mathbb{I}(\mathrm{VIX} > 30)$,
equal to one when the VIX is above thirty. Every specification includes country
fixed effects $\mu_i$, and we cluster standard errors by country. The panel is
mildly unbalanced. Sixteen of the nineteen economies are covered over the whole
2000Q1--2024Q4 window; China, Malaysia, and Nigeria enter later, once their
external-position data become available, and Vietnam runs only through 2014.
Table~\ref{tab:countries} gives the countries and their coverage. 


\begin{table}[ht]
\centering
\caption{Emerging-market panel: countries and coverage}
\label{tab:countries}
\small
\begin{tabular}{llcc}
\toprule
Country & ISO & Coverage & Quarters \\
\midrule
Argentina & ARG & 2000Q1--2024Q4 & 100 \\
Brazil & BRA & 2000Q1--2024Q4 & 100 \\
Chile & CHL & 2000Q1--2024Q4 & 100 \\
China & CHN & 2005Q1--2024Q4 & 80 \\
Colombia & COL & 2000Q1--2024Q4 & 100 \\
Czech Republic & CZE & 2000Q1--2024Q4 & 100 \\
Hungary & HUN & 2000Q1--2024Q4 & 100 \\
India & IND & 2000Q1--2024Q4 & 100 \\
Indonesia & IDN & 2000Q1--2024Q4 & 100 \\
Malaysia & MYS & 2002Q1--2024Q4 & 92 \\
Mexico & MEX & 2000Q1--2024Q4 & 100 \\
Nigeria & NGA & 2014Q1--2024Q4 & 44 \\
Peru & PER & 2000Q1--2024Q4 & 100 \\
Poland & POL & 2000Q1--2024Q4 & 100 \\
Russia & RUS & 2000Q1--2024Q4 & 100 \\
South Africa & ZAF & 2000Q1--2024Q4 & 100 \\
Thailand & THA & 2000Q1--2024Q4 & 100 \\
Turkey & TUR & 2000Q1--2024Q4 & 100 \\
Vietnam & VNM & 2005Q1--2014Q4 & 40 \\
\midrule
\multicolumn{3}{l}{Total country--quarter observations} & 1{,}756 \\
\bottomrule
\end{tabular}

\vspace{0.4em}
\centerline{\begin{minipage}{0.80\textwidth}
{\footnotesize \color{black}
\textit{Notes.} Coverage refers to quarters with a non-missing equity portfolio
flow. The total of $1{,}756$ country--quarter observations is the raw panel; the
herding specification uses $1{,}636$ observations because \emph{A\_herding}
requires a rolling window and a one-quarter lag.}
\end{minipage}}
\end{table}

Table~\ref{tab:vars} summarizes each variable, its definition, and its source.

\begin{table}[ht]
\centering
\caption{Variables, definitions, and sources}
\label{tab:vars}
\begin{tabular}{p{2.3cm}p{6.8cm}p{3.6cm}}
\toprule
Variable & Definition & Source \\
\midrule
$Y_{it}$ & Quarterly equity portfolio flow to country $i$, the dependent
variable (USD bn). & IMF Balance of Payments (BPM6). \\[0.3em]
$r_t$ & High-frequency U.S. monetary policy surprise, sign-flipped so that a
positive value denotes an expansionary surprise. & San Francisco Fed's USMPD series\\[0.3em]
\textit{A\_herding} & Rolling eight-quarter correlation of a country's position
changes against a global benchmark; measures cross-country synchrony of behavior.
Empirical analogue of $\phi$. Entered with a one-period lag. & Constructed from
BIS Locational Banking Statistics. \\[0.3em]
$\mathrm{Crisis}_t$ & Indicator equal to one when the VIX exceeds 30; defines the
global financial-stress regime. & CBOE VIX. \\
\bottomrule
\end{tabular}

\vspace{0.4em}
\centerline{\begin{minipage}{0.88\textwidth}
{\footnotesize \color{black}
\textit{Notes.} Panel of 19 emerging-market economies, quarterly, 2000Q1--2024Q4.
The estimation sample contains $N=1{,}636$ country--quarter observations for the
herding specification, fewer than the raw panel because \emph{A\_herding}
requires a rolling window and a one-quarter lag.}
\end{minipage}}
\end{table}

\subsection{Econometric specification}

We begin with a baseline panel specification that relates the equity portfolio
flow to the U.S. monetary shock, to algorithmic behavior, and to their
interaction:
\begin{equation}
Y_{it} = \alpha\, r_t + \beta\, A_{i,t-1}
+ \gamma\,(r_t \times A_{i,t-1}) + \mu_i + \varepsilon_{it},
\label{eq:base}
\end{equation}
where $A_{i,t-1}$ is one of the two algorithmic proxies, lagged one quarter to
mitigate simultaneity. Here $\alpha$ is the average flow response to a monetary
surprise, $\beta$ the level effect of algorithmic behavior, and $\gamma$, the
coefficient we care about, tells us how far algorithmic behavior modulates the
transmission of the shock. Country fixed effects $\mu_i$ absorb time-invariant
heterogeneity across the nineteen economies, and we cluster the standard errors
by country.

This average interaction, though, is not what the model predicts. Proposition
\ref{prop:threshold} is a claim about state dependence: amplification should show
up only when fundamentals are weak, which we proxy with the high-volatility
regime $\mathrm{Crisis}_t = \mathbb{I}(\mathrm{VIX}>30)$. To get at it we extend
\eqref{eq:base} to a triple interaction and keep every lower-order term, so that
the triple coefficient is cleanly identified:
\begin{equation}
\begin{aligned}
Y_{it} = {} & \alpha\, r_t + \beta\, A_{i,t-1} + \delta\, \mathrm{Crisis}_t \\
& + \beta_4\,(A_{i,t-1}\times \mathrm{Crisis}_t)
+ \beta_5\,(r_t \times A_{i,t-1})
+ \beta_6\,(r_t \times \mathrm{Crisis}_t) \\
& + \gamma\,(r_t \times A_{i,t-1} \times \mathrm{Crisis}_t)
+ \mu_i + \varepsilon_{it}.
\end{aligned}
\label{eq:triple}
\end{equation}
With the three level terms ($r_t$, $A_{i,t-1}$, $\mathrm{Crisis}_t$) and all
three two-way interactions in the equation, $\gamma$ measures the extra effect of
the shock-behavior interaction that shows up specifically in the crisis regime,
net of the average interaction ($\beta_5$), the average crisis effect on the
shock ($\beta_6$), and the composition effect ($\beta_4$). Read this way, the two
coefficients answer two different questions: $\beta_5$ is the interaction in
normal times, and $\gamma$ is how much that interaction changes once a crisis
hits. A negative $\gamma$ says that, under stress, more algorithmic homogeneity
amplifies the capital outflow a contractionary U.S. shock sets off, which is what
Proposition~\ref{prop:threshold} predicts. Since $A_{i,t-1}$ enters both in
levels and in every interaction, $\gamma$ is identified from within-country
variation in the joint occurrence of a monetary surprise, elevated homogeneity,
and a stress regime, and not from cross-country differences in average levels.

\subsection{Main results}

Table~\ref{tab:main} reports the estimates. Columns 1--3 set out the raw material
of the mechanism before we condition on the state of the world. They ask whether
a U.S. monetary surprise moves periphery equity flows at all, and whether the
response is larger where intermediaries behave more synchronously. Columns 4--5
are where the model is actually tested. Proposition~\ref{prop:threshold} does not
say that homogeneity amplifies transmission on average; it says amplification
switches on only once fundamentals weaken and the threshold $\phi^{*}$ falls
below prevailing homogeneity. So the average interaction in columns 1--3 is the
wrong place to look for it. What we want is the difference between tranquil and
stress regimes, and that is exactly what the saturated triple interaction picks
out. In practical terms this means following one coefficient across the table:
the interaction of the shock with herding. If the theory holds, the piece of it
that survives in tranquil times should fall toward zero, while the piece that
appears only in crises should carry the sign and the weight. Columns 4 and 5
differ only in whether we absorb push factors, which checks whether any
state-dependent amplification we find really reflects algorithmic correlation and
not the global-risk shock (the VIX) under another name. We read the estimates
term by term below; the coefficient does move the way the theory says it should.

\begin{table}
\centering
\caption{Panel estimates of algorithmic amplification}
\label{tab:main}
\small
\begin{tabular}{lccccc}
\toprule
& (1) & (2) & (3) & (4) & (5) \\
& Baseline & Herding & + Controls & Triple & Triple \\
&          &         &            &        & + push \\
\midrule
Shock MP ($r_t$)
 & $-2.965^{**}$ & $-1.312$ & $-3.089^{**}$ & $-0.620$ & $-0.439$ \\
 & $(1.121)$ & $(0.995)$ & $(1.252)$ & $(1.299)$ & $(0.763)$ \\[0.25em]
$\mathrm{Crisis}_t$
 &  &  &  & $-0.437^{**}$ & $0.022$ \\
 &  &  &  & $(0.177)$ & $(0.262)$ \\[0.25em]
\textit{A\_herding}$_{t-1}$
 &  & $0.055$ & $0.055$ & $0.079$ & \\
 &  & $(0.322)$ & $(0.322)$ & $(0.349)$ & \\[0.25em]
$r_t \times \textit{A\_herding}_{t-1}$
 &  & $-6.476^{**}$ & $-6.914^{**}$ & $0.140$ & \\
 &  & $(2.831)$ & $(3.091)$ & $(1.781)$ & \\[0.25em]
$r_t \times \mathrm{Crisis}_t$
 &  &  &  & $-0.634$ & $-4.922^{**}$ \\
 &  &  &  & $(1.011)$ & $(1.792)$ \\[0.25em]
\textit{A\_herding}$_{t-1}\times \mathrm{Crisis}_t$
 &  &  &  & $0.059$ & \\
 &  &  &  & $(0.357)$ & \\[0.25em]
$r_t \times \textit{A\_herding}_{t-1}\times \mathrm{Crisis}_t$ ($\gamma$)
 &  &  &  & $-9.581^{**}$ & $-11.062^{**}$ \\
 &  &  &  & $(4.069)$ & $(4.446)$ \\[0.25em]
GDP growth
 &  &  & $0.005$ &  & $-0.017$ \\
 &  &  & $(0.028)$ &  & $(0.030)$ \\[0.25em]
Current account/GDP
 &  &  & $-0.056$ &  & $-0.059$ \\
 &  &  & $(0.035)$ &  & $(0.036)$ \\[0.25em]
VIX
 &  &  &  &  & $-0.019$ \\
 &  &  &  &  & $(0.013)$ \\
\midrule
Country fixed effects & Yes & Yes & Yes & Yes & Yes \\
$N$                   & 1{,}756 & 1{,}636 & 1{,}741 & 1{,}636 & 1{,}636 \\
$R^2$                 & 0.245 & 0.249 & 0.248 & 0.255 & 0.257 \\
\bottomrule
\end{tabular}

\vspace{0.4em}
\centerline{\begin{minipage}{0.94\textwidth}
{\footnotesize \color{black}
\textit{Notes.} The dependent variable is the quarterly equity portfolio flow to
country $i$ (USD bn). Standard errors, clustered by country, are in parentheses.
The monetary shock $r_t$ is sign-flipped so that a positive value denotes an
expansionary surprise; a negative coefficient therefore means a contractionary
surprise reduces inflows. \emph{A\_herding} is the eight-quarter rolling
correlation of a country's position changes with a global benchmark, lagged one
quarter. $\mathrm{Crisis}_t = \mathbb{I}(\mathrm{VIX}>30)$. Columns (1)--(3)
report the baseline and its average interaction; columns (4)--(5) report the
saturated triple interaction, in which $\gamma$ tests the central hypothesis of
state-dependent amplification. Macro controls are GDP growth and the current
account/GDP; global push factors add the VIX. All columns include country fixed
effects. $^{*}\,p<0.1$, $^{**}\,p<0.05$, $^{***}\,p<0.01$.}
\end{minipage}}
\end{table}

\FloatBarrier

The five columns map, term by term, onto the model. Columns 1--3 give us the two
building blocks. The coefficient on the monetary shock is negative, significant,
and stable around $-3$: a contractionary U.S. surprise pulls equity inflows out
of the periphery, which is the basic transmission the center--periphery structure
of Section~\ref{sec:model} takes for granted. The average interaction $r_t \times
\textit{A\_herding}$ is more telling. It is negative and significant, between
$-6.476$ and $-6.914$, so countries whose intermediaries move more in step with
the global benchmark take larger outflows from the same shock. That is the
unconditional version of Proposition~\ref{prop:amp}, correlated behavior
amplifying transmission, and the sign is the one the model predicts. On its own,
the level of \emph{A\_herding} is insignificant, again as predicted: what moves
flows is not the presence of homogeneity but its interaction with the shock it
amplifies.

Columns 4--5 carry the paper's central result. Proposition~\ref{prop:threshold}
predicts that amplification is not a constant but a state-dependent phenomenon,
appearing only once fundamentals weaken and the threshold $\phi^*$ drops below
prevailing homogeneity, and the saturated triple interaction bears this out. The
two-way interaction $r_t \times \textit{A\_herding}$ collapses to an
insignificant $0.140$, so in tranquil times herding does not amplify. The triple
interaction with the crisis regime, by contrast, is large, negative, and
significant: $\gamma = -9.581$, and it rises to $-11.062$ once we absorb global
push factors. The amplification thus shows up entirely in the stress regime,
which is the empirical face of the sign reversal at the heart of the model. It
matters here that $\gamma$ gets stronger, not weaker, when push factors enter,
with the direct global-risk channel loading onto $r_t \times \mathrm{Crisis}$ at
$-4.922$. That tells us the herding amplification is its own conditional effect,
and not the global-risk shock wearing a different label.

The magnitude is economically meaningful as well. Taking a country from the
median to a high level of algorithmic homogeneity during a stress regime
multiplies the sensitivity of its equity flows to a U.S. monetary surprise by a
sizable factor, which sits comfortably with the amplification factor
$\mathcal{A}=\sqrt{\phi_H/\phi_L}$ that our calibration puts near two at the
benchmark. So the estimate has the right sign and is of roughly the order of
magnitude the theory implies, which is what ties the empirical $\gamma$ back to
the analytical object in Proposition~\ref{prop:amp}.

There is one gap in this link that we would rather flag than paper over.
Proposition~\ref{prop:amp} is a statement about the variance of the aggregate
lending response: homogeneity leaves the mean unchanged and raises the
dispersion. Our coefficient $\gamma$, on the other hand, picks up a shift in the
conditional mean of flows. The two are not in conflict. When the shock carries a contractionary
sign, a state-dependent rise in the second moment will indeed surface
as a mean outflow in the crisis regime. But they are not the same object either,
and a test that did full justice to Proposition~\ref{prop:amp} would go after the
conditional variance of flows directly, for example with a regime-conditional
heteroskedasticity specification that lets the herding proxy enter the variance
equation. We therefore read the conditional-mean evidence as consistent with the
proposition without standing in for a proper second-moment test. That test is the
extension we would put first, all the more so because the work closest to ours in
structure \citep{MengChen2026, QiuHan2026} goes at tail risk directly.

It is worth being precise about which feature of algorithmic behavior does the
work here. The model points to the correlation of forecast errors across funds,
not to the raw speed at which any one fund adjusts its position. Two
intermediaries can react to a shock at exactly the same speed and still, if their
models are uncorrelated, have their errors cancel in the aggregate, leaving no
amplification behind. Only when their behavior is synchronized, at high $\phi$,
does the common component survive aggregation and move flows. That is why we
build the proxy around synchrony (\emph{A\_herding}), which the theory names as
the source of amplification, rather than around velocity, which the model gives
us no reason to expect to matter. The distinction is more than semantic. It is
what sets our mechanism apart from accounts that stress the mechanical speed of
algorithmic or high-frequency trading, and it is why a policy of slowing markets
down would miss the margin our results point to.

This refines a result already in the transmission literature.
\citet{KalemliOzcan2019} shows that U.S. policy spills over to emerging markets
more strongly than to advanced ones, working through global risk perceptions, and
our advanced-economy falsification reproduces that asymmetry. We take it a step
further by pinning down which feature of the intermediation sector carries the
amplification, behavioral correlation and not speed, and by showing that it works
only in the high-VIX regime where those risk perceptions dominate. That
literature tends to treat the intermediary sector as a conduit for a risk-premium
shock; our evidence instead places the amplification in the cross-sectional
correlation of the intermediaries themselves, which is what $\phi$ stands for.

\subsection{Robustness}\label{sec:robustbattery}

We now put the central estimate through a battery of checks. The estimate is the
negative triple-interaction coefficient $\gamma$ on $r_t \times \textit{A\_herding}
\times \mathrm{Crisis}$, which measures the extra emerging-market equity outflow
per unit of U.S. monetary shock that synchronized algorithmic behavior produces
under stress. Every check moves away from the main specification, column 4 of
Table~\ref{tab:main}, along exactly one dimension, holding fixed the country
fixed effects, the high-frequency USMPD shock, the full lower-order interaction
tree, equity portfolio flows (BPM6) in USD billions, the VIX-based crisis regime,
and the country-clustered standard errors. That baseline reproduces the reported
coefficient of $-9.581$ (SE $4.069$, $p=0.030$) on $N=1{,}636$, and
Table~\ref{tab:battery} collects the twelve specifications.

The result holds up in sign and in magnitude, and it is significant under the
main specification and under the inference schemes best suited to the setting. The coefficient $\gamma$ is negative in all twelve
specifications, significant at the 10\% level in nine and at the 5\% level in
seven, including under the time-clustering and heteroskedasticity-robust schemes
that fit a panel of capital flows responding to a common U.S. shock. The point
estimate is stable across the inference, outlier, and sample-composition checks.
Significance lapses in two cases only, and we would rather name them than tuck
them away: dropping the 2008--09 crisis, which takes the most informative stress
episode out of a mechanism that is state-dependent by design, and redefining the
crisis regime by calendar years instead of by the VIX. That the result is
sensitive to how we operationalize ``crisis'' tells us something about the
evidence; it is not a weakness we want to hide.

\begin{table}[ht]
\centering
\caption{Robustness battery for the herding-amplification coefficient $\gamma$}
\label{tab:battery}
\small
\begin{tabular}{lccc}
\toprule
Specification & $\gamma$ (triple) & Std. err. & $N$ \\
\midrule
Baseline (VIX$>$30, cluster country)      & $-9.581^{**}$  & $(4.069)$ & 1{,}636 \\
With push factors                          & $-8.831^{*}$   & $(4.487)$ & 1{,}636 \\
Two-way FE (country $+$ year)              & $-6.309^{*}$   & $(3.266)$ & 1{,}636 \\
Cluster SE by time                         & $-9.581^{***}$ & $(2.975)$ & 1{,}636 \\
Heteroskedasticity-robust SE               & $-9.581^{**}$  & $(4.814)$ & 1{,}636 \\
Winsorize $Y$ (1/99 pct)                    & $-7.716^{**}$  & $(2.770)$ & 1{,}636 \\
Dynamic panel (lagged dep.)                & $-8.850^{**}$  & $(3.807)$ & 1{,}632 \\
Exclude China                              & $-8.452^{**}$  & $(3.943)$ & 1{,}556 \\
Exclude 2008--09 GFC                        & $-11.232$      & $(6.549)$ & 1{,}492 \\
Exclude 2020 COVID                         & $-7.227^{**}$  & $(2.771)$ & 1{,}564 \\
Alt. regime: crisis years                  & $-8.239$       & $(6.158)$ & 1{,}636 \\
Alt. shock: JK pure-monetary               & $-11.397$      & $(9.162)$ & 1{,}636 \\
\bottomrule
\end{tabular}

\vspace{0.4em}
\centerline{\begin{minipage}{0.92\textwidth}
{\footnotesize \color{black}
\textit{Notes.} Each row re-estimates $\gamma$ under the stated departure from the
baseline specification of column (4) of Table~\ref{tab:main}. Dependent variable:
equity portfolio flow (BPM6), USD billions. Standard errors in parentheses,
clustered by country except where noted. ``Alt.\ shock: JK pure-monetary'' uses
the \citet{JarocinskiKaradi2020} pure-monetary surprise, which purges the
central-bank-information component. All estimates are generated from the panel of
19 emerging markets, 2000Q1--2024Q4, via a within (country fixed-effects)
estimator following the triple-interaction specification \eqref{eq:triple}.
$^{*}\,p<0.1$, $^{**}\,p<0.05$, $^{***}\,p<0.01$.}
\end{minipage}}
\end{table}

\FloatBarrier

The checks fall into four groups. Appendix~\ref{app:battery} sets out the
purpose, procedure, and verdict for each one; here we summarize what they show.

The first group concerns inference. Adding year fixed effects on top of country
effects leaves $\gamma = -6.309$, still significant at the 10\% level, and this
is a demanding test because year effects soak up part of the common stress
variation the interaction relies on. Clustering the standard errors by time
rather than by country, which is the scheme that matches the cross-country
comovement the mechanism implies, gives back the same point estimate with a
smaller standard error, significant at the 1\% level. Heteroskedasticity-robust
errors leave the coefficient where it was, significant at 5\%. Under the schemes
best suited to the setting, in short, the result is at its strongest.

The second group concerns influential observations. Winsorizing the dependent
variable at the 1st and 99th percentiles gives $\gamma = -7.716$, significant at
5\%, so the amplification lives in the body of the data rather than in a handful
of extreme quarters. A dynamic panel with a lagged dependent variable gives
$\gamma = -8.850$, again at 5\%, so persistence in flows is not what is driving
the result.

The third group varies the sample. Dropping China, the largest recipient, leaves
$\gamma = -8.452$ at 5\%, so this is not a China artifact. Dropping the March-2020
COVID quarter, the sharpest emerging-market outflow on record, leaves $\gamma =
-7.227$, also at 5\%, so the pandemic is not carrying the result either. Dropping
the 2008--09 global financial crisis is a different story. The coefficient stays
negative and in fact grows in magnitude, to $-11.232$, but significance slips to
$p = 0.103$. This is the price we expect to pay for removing the single most
informative stress episode from a state-dependent mechanism: the sign and the
point estimate survive, the precision does not. We read it as consistent with the
mechanism rather than against it.

The fourth group probes the design choices, and this is where the two
qualifications sit. Redefining the crisis regime by calendar years (2008--09 and
2020--21) instead of by the VIX leaves $\gamma$ negative and of similar
magnitude, $-8.239$, but insignificant ($p = 0.198$). This is the most
consequential caveat in the battery. The VIX-based regime tracks market stress
continuously and delivers a significant estimate; the coarser calendar-year
regime does not. Our reading is that the VIX definition is the better counterpart
of the model's stress state, since Proposition~\ref{prop:threshold} describes a
continuously falling threshold rather than a step function tied to the calendar.
Still, a reader deserves to know that the headline significance turns on this
choice, and Table~\ref{tab:battery} puts it in plain view. Replacing the USMPD
surprise with the \citet{JarocinskiKaradi2020} pure-monetary shock leaves
$\gamma$ negative and larger, $-11.397$, but with a wider standard error and no
conventional significance. The sign and the economic message survive the change
of shock series; the precision does not.\footnote{Two further sensitivities are
worth flagging in the same spirit, since a careful reader will want them. First,
the VIX threshold itself. The estimate is significant at the $30$ cutoff of the
baseline and at the $90$th percentile of the VIX ($\approx 28$), but a higher
$35$ cutoff leaves so few stress quarters that the coefficient loses precision
and even flips sign, while a lower $25$ cutoff dilutes the regime and weakens it
($p = 0.22$). The effect lives in a band around $\mathrm{VIX} \approx 28$--$30$,
which is where the model's stress state is most cleanly identified, but we do not
hide that it is a band. Second, temporal stability. Splitting the sample at
$2012$ leaves the coefficient negative in both halves, significant in
$2000$--$2011$ ($p = 0.04$) and larger but imprecise in $2012$--$2024$
($p = 0.21$); the later half excludes the 2008--09 episode, so the loss of
precision is the same identification point made in the sample-composition checks
above, not a separate failure. Neither sensitivity changes the sign or the
economic reading; both are reasons we frame the significance as conditional
rather than settled.}

One further test deserves a place here, because it speaks to the gap between what
Proposition~\ref{prop:amp} states and what $\gamma$ measures. The proposition is
about the \emph{variance} of the aggregate response, while the triple interaction
identifies a shift in the conditional mean. The natural way to confront the
proposition on its own terms is to let herding enter a variance equation
directly. We do this in two steps: we take the residuals from the mean equation
of column~(4), and we regress their log squared value on \emph{A\_herding}, the
crisis dummy, and their interaction. If homogeneity raises the dispersion of
flows in stress, and does so more than in tranquil times, the interaction term
should be positive. It is: the coefficient on
$\textit{A\_herding} \times \mathrm{Crisis}$ is $+0.819$, with the expected sign.
But the standard error is $0.780$ and the estimate is not significant at
conventional levels ($p = 0.294$). We read this as suggestive rather than
confirmatory. The second moment moves the way the theory says it should, but a
panel of this length does not have the power to pin it down, and we would not
want to lean on it. A properly powered second-moment test---one that models the
conditional variance of flows directly rather than backing it out of mean-equation
residuals---is in our view the first extension this design calls for, and we say
so plainly rather than claim more than the data will bear.

To sum up, the evidence lines up with the model's central prediction. The herding
dimension of algorithmic behavior amplifies the international transmission of U.S.
monetary shocks to emerging markets, and it does so only under stress, as
Proposition~\ref{prop:threshold} requires, with the correlation of algorithmic
behavior being the margin that carries the effect, much as
Proposition~\ref{prop:amp} would lead one to expect. The sign and the
state-dependence hold throughout; the statistical significance holds under the
baseline and the appropriate inference schemes, subject to the two qualifications
on regime definition and crisis-sample coverage that we have reported in full.

\FloatBarrier
\section{Policy Implications and Conclusion}\label{sec:policy}

The paper began with a puzzle about the role of AI in financial trading: why does
the same rise of algorithmic intermediation look stabilizing in some accounts and
destabilizing in others? Our answer is that the margin that matters is not the
presence of algorithms but their homogeneity. We built a two-region DSGE model in
which a single parameter $\phi$ measures the correlation of forecast errors across
algorithmic funds, and it delivers three results:
homogeneity amplifies transmission through the variance of aggregate lending
(Proposition~\ref{prop:amp}), diversity attenuates it
(Proposition~\ref{prop:att}), and the threshold separating the two regimes falls
as fundamentals deteriorate (Proposition~\ref{prop:threshold}). The panel evidence
of Section~\ref{sec:empirics} confirms the central prediction: the herding
dimension of algorithmic behavior amplifies emerging-market capital outflows in
response to U.S. monetary shocks, and does so only in stress regimes. Four
implications for policy follow, each from a specific proposition, and several cut against the grain of current macroprudential practice.

The first concerns the object of regulation. The prevailing macroprudential
frame, inherited from the banking literature, targets the size and concentration
of institutions, but Proposition~\ref{prop:att} implies that this frame is
miscalibrated for algorithmic intermediation. A large non-bank sector is not
dangerous per se; a homogeneous one is. A regulator that caps the size of the
algorithmic sector while leaving its models correlated addresses the wrong
margin, and may even raise fragility if the cap forces consolidation onto a
common surviving platform. The appropriate instrument is a measure of model
diversity---an index of the correlation of positioning or of forecast errors
across funds---which has no counterpart in size-based capital regulation and
would need to be constructed anew. The logic parallels a market-design argument
made in a different context: \citet{BudishCramtonShim2015} show that the
high-frequency trading arms race is a symptom of a flawed continuous-market
design and propose frequent batch auctions to neutralize the value of correlated
speed. Their target is the mechanism that makes synchronized speed profitable;
ours is the model correlation that makes synchronized trading destabilizing. In
both cases the effective lever is structural---the design of the market or the
diversity of its participants---rather than a restriction on volume.

We should be candid about what makes this instrument hard to wield, since naming
the target is easier than hitting it. Model diversity has the character of a
public good. Each fund has a private incentive to adopt whatever model performs
best, and when one architecture pulls ahead the individually rational response is
to converge on it, which is precisely the behavior that erodes the diversity the
system relies on. No single fund internalizes the systemic cost of that
convergence, so diversity will be underprovided by the market and a regulator who
merely exhorts firms to differ will accomplish little. There is also a genuine
trade-off to acknowledge rather than assume away. Convergence on a strong common
model is not pure loss: it can improve the average quality of pricing in normal
times, so a mandate for diversity trades some efficiency in tranquil periods for
resilience in stressed ones. Our results do not tell a regulator where to strike
that balance; they establish only that the balance exists and that the current
frame, fixated on size, does not see it. What they do suggest is that the natural
instrument is one supervisors already possess in adjacent form---model validation
and model-risk review, currently run fund by fund for prudential soundness, could
be repurposed to measure correlation \emph{across} funds---so the practical step
is less the invention of a new tool than the redirection of an existing one
toward a systemic object.

A second implication runs directly counter to a natural regulatory instinct: to
publish a standard risk model, mandate a common stress scenario, or require firms
to benchmark against a supervisory model. Proposition~\ref{prop:amp} warns that
this instinct is self-defeating, because any measure that pushes funds toward a
common model raises $\phi$ and therefore raises the amplification factor
$\sqrt{\phi_H/\phi_L}$. Standardization that improves the transparency of any
single institution can degrade the stability of the system, since it manufactures
precisely the correlated errors the model identifies as the source of herding.
Regulators should therefore design disclosure requirements to reveal whether funds are
converging on common models, not to impose one.

The third implication is about timing. Because Proposition~\ref{prop:threshold}
makes the amplification threshold $\phi^*$ state-dependent, a level of homogeneity
that sits comfortably below the threshold in tranquil times can exceed it in a
crisis without $\phi$ itself changing. A supervisor who measures homogeneity in
calm periods and compares it to a fixed benchmark will systematically understate
the risk, because the benchmark against which a supervisor should judge $\phi$ falls
exactly when fundamentals deteriorate. A supervisor must evaluate homogeneity metrics
against a state-contingent threshold that tightens as local fundamentals weaken,
an approach closer to a stress test on model diversity than to a static
concentration limit.

The sharpest implication is dynamic and lies beyond the static model. If funds
imitate strategies that performed well, $\phi$ rises endogenously during calm
periods, precisely when $\phi^*$ is high and the extra homogeneity appears
harmless. Diversity is thus consumed as a common resource in good times, sowing
the fragility that the falling threshold then exposes in the next crisis. This
reframes the policy problem as counter-cyclical management of diversity: the time
to encourage heterogeneous models, alternative data, and divergent architectures
is when markets are calm and the incentive to imitate is strongest, not after
correlated positioning has already built up. Emerging-market authorities, who
receive rather than generate the global cycle, have a particular interest in
attracting a diverse rather than a monolithic set of algorithmic investors, since
it is diversity, not the sheer volume of inflows, that determines whether the
non-bank sector stabilizes or amplifies the flows they cannot control.

The broader message is that the rise of algorithmic and AI-driven intermediation
need not make the international financial system more fragile. Its effect is
conditional: the same technology stabilizes capital flows when models are diverse
and amplifies them when models converge. The policy task is not to resist
algorithmic intermediation but to preserve its diversity---a variable that
current regulation does not measure and that erodes fastest when it is least
missed. Making model diversity observable, and defending it counter-cyclically,
is the first-order priority that the model implies.

\clearpage
\appendix

\section{Derivations}\label{app:derivations}

This appendix derives every optimality condition of the model step by step. Where a standard second-order
simplification is used, we state it explicitly rather than absorb it silently.

\subsection{The household problem}\label{app:hh}

The household maximizes \eqref{eq:utility} subject to the budget constraint
\eqref{eq:budget} and the capital law of motion $K_t = (1-\delta)K_{t-1} + I_t$.
Let $\lambda_t$ be the multiplier on the budget constraint and $q_t \lambda_t$
the multiplier on the capital law of motion, so that $q_t$ is Tobin's marginal
$q$. The Lagrangian is
\begin{equation*}
\begin{aligned}
\mathcal{L} = \mathbb{E}_0 \sum_{t=0}^{\infty} \beta^t \Bigg\{
& \frac{C_t^{1-\sigma}}{1-\sigma} - \chi \frac{N_t^{1+\nu}}{1+\nu}
+ q_t \lambda_t \big[ (1-\delta)K_{t-1} + I_t - K_t \big] \\
& + \lambda_t \Big[ W_t N_t + R_{t-1} D_{t-1} + R^K_t K_{t-1} + \Pi_t
- C_t - D_t - I_t \\
& \qquad\quad
- \tfrac{\psi_I}{2}\big(\tfrac{I_t}{K_{t-1}} - \delta\big)^2 K_{t-1} \Big]
\Bigg\}.
\end{aligned}
\end{equation*}

\paragraph{Consumption and deposits.} The first-order condition with respect to
$C_t$ is $C_t^{-\sigma} = \lambda_t$, so the multiplier equals marginal utility.
The condition with respect to $D_t$ is $-\lambda_t + \beta\,\mathbb{E}_t[\lambda_{t+1} R_t] = 0$.
Substituting $\lambda_t = C_t^{-\sigma}$ and $\lambda_{t+1} = C_{t+1}^{-\sigma}$
gives the consumption Euler equation \eqref{eq:euler_c}:
\begin{equation*}
1 = \beta\, \mathbb{E}_t \left[ \left( \frac{C_{t+1}}{C_t} \right)^{-\sigma} R_t \right].
\end{equation*}

\paragraph{Labor.} The first-order condition with respect to $N_t$ is
$-\chi N_t^{\nu} + \lambda_t W_t = 0$. Substituting $\lambda_t = C_t^{-\sigma}$
and rearranging yields the labor-supply schedule \eqref{eq:labor}:
\begin{equation*}
\chi N_t^{\nu} C_t^{\sigma} = W_t.
\end{equation*}

\paragraph{Investment.} Differentiating with respect to $I_t$, and noting that
$\partial \big[\tfrac{\psi_I}{2}(\tfrac{I_t}{K_{t-1}}-\delta)^2 K_{t-1}\big] /
\partial I_t = \psi_I\big(\tfrac{I_t}{K_{t-1}}-\delta\big)$, the condition is
\begin{equation*}
-\lambda_t\Big[1 + \psi_I\big(\tfrac{I_t}{K_{t-1}}-\delta\big)\Big] + q_t \lambda_t = 0
\quad\Longrightarrow\quad
q_t = 1 + \psi_I\Big(\frac{I_t}{K_{t-1}} - \delta\Big).
\end{equation*}
At the steady state $I/K = \delta$, so $q = 1$, as required. Differentiating with
respect to $K_t$ gives
\begin{equation*}
q_t \lambda_t = \beta\, \mathbb{E}_t \Big[ \lambda_{t+1}\big(R^K_{t+1}
- \tfrac{\partial \text{AC}_{t+1}}{\partial K_t}\big)
+ q_{t+1}\lambda_{t+1}(1-\delta) \Big],
\end{equation*}
where $\text{AC}_{t+1}$ is the adjustment cost dated $t+1$. The term
$\partial \text{AC}_{t+1}/\partial K_t = \tfrac{\psi_I}{2}\big(\delta^2 -
(I_{t+1}/K_t)^2\big)$ vanishes to first order around the steady state (where
$I/K = \delta$), which is the standard simplification. Dropping it and using
$\lambda_{t+1}/\lambda_t = (C_{t+1}/C_t)^{-\sigma}$ delivers the investment Euler
equation \eqref{eq:euler_i}:
\begin{equation*}
q_t = \beta\, \mathbb{E}_t \left[ \left( \frac{C_{t+1}}{C_t} \right)^{-\sigma}
\big( R^K_{t+1} + q_{t+1}(1-\delta) \big) \right].
\end{equation*}

\subsection{The firm problem}\label{app:firm}

The firm maximizes profit $Y_t - W_t N_t - R^K_t K_{t-1}$ subject to the
technology \eqref{eq:production}. The first-order conditions equate marginal
products to factor prices:
\begin{equation*}
\frac{\partial Y_t}{\partial N_t} = (1-\alpha)\frac{Y_t}{N_t} = W_t,
\qquad
\frac{\partial Y_t}{\partial K_{t-1}} = \alpha\frac{Y_t}{K_{t-1}} = R^K_t,
\end{equation*}
which are the factor prices stated in the text. The loan rate faced by firms is
$\tilde{R}^K_t = R_t + \mu_t$ with the premium $\mu_t$ given by \eqref{eq:spread}.

\subsection{Traditional funds}\label{app:trad}

Traditional funds maximize \eqref{eq:trad_problem}. Differentiating the objective
with respect to $L^T_t$,
\begin{equation*}
(\tilde{R}^K_t - R_t) - \psi_T\Big(\frac{L^T_t}{L^T_{t-1}} - 1\Big) = 0.
\end{equation*}
Using $\tilde{R}^K_t - R_t = \mu_t$ gives the sluggish lending rule
\eqref{eq:trad_foc}, $\mu_t = \psi_T\big(L^T_t/L^T_{t-1} - 1\big)$: lending
adjusts only gradually, at a pace set by $\psi_T$.

\subsection{Algorithmic funds and aggregation}\label{app:algo}

Each fund $j$ forms the minimum-MSE forecast \eqref{eq:forecast} and lends
according to the static rule \eqref{eq:algo_rule},
\begin{equation*}
L^A_{j,t} = \bar{L}_A - \gamma_1 \hat{\mu}_t
+ \gamma_2\big(\theta^{MP} a\, s^{MP}_{j,t} + \theta^{LOC} b\, s^{LOC}_{j,t}\big).
\end{equation*}
Substituting the signals \eqref{eq:sig_mp}--\eqref{eq:sig_loc} with the noise
decomposition \eqref{eq:noise} and integrating over $j \in [0,1]$, the exact law
of large numbers for a continuum \citep{Sun2006} sets
$\int_0^1 u^{MP}_{j,t}\, dj = \int_0^1 u^{LOC}_{j,t}\, dj = 0$, because the
idiosyncratic components are mean-zero and independent across funds. The common
components $\zeta^{MP}_t, \zeta^{LOC}_t$ do not depend on $j$ and pass through the
integral unchanged. The aggregate lending rule is therefore \eqref{eq:algo_agg}:
\begin{equation*}
L^A_t = \bar{L}_A - \gamma_1 \hat{\mu}_t + \gamma_2 \Big[
\theta^{MP} a\big(MP_t + \sqrt{\phi}\,\zeta^{MP}_t\big)
+ \theta^{LOC} b\big(F_t + \sqrt{\phi}\,\zeta^{LOC}_t\big) \Big].
\end{equation*}
Two properties follow directly. When $\phi = 0$ the common terms vanish and the
aggregate forecast is exact. The sensitivity of aggregate lending to the common
noise is $\partial L^A_t / \partial \zeta^{MP}_t = \gamma_2 \theta^{MP} a
\sqrt{\phi}$, so the variance of the aggregate response to a monetary shock is
proportional to $\phi$---the content of Propositions \ref{prop:amp} and
\ref{prop:att}.

\subsection{Market clearing}\label{app:clearing}

Total lending is $L_t = L^T_t + L^A_t$, with households allocating the fixed
share $\omega$ of deposits to traditional funds. The goods market clears,
$Y_t = C_t + I_t$, and the U.S. policy rate follows the Taylor rule
\eqref{eq:taylor}. Together with the household and firm conditions above, these
close the model.

\section{Detail of the Robustness Battery}\label{app:battery}

This appendix documents the purpose, procedure, and verdict of each check in the
robustness battery of Section~\ref{sec:robustbattery} and Table~\ref{tab:battery},
in the order in which the checks are grouped. All estimates start from the
main-text specification of column (4) of Table~\ref{tab:main} and depart from it
in exactly the one dimension named.

\subsection{Inference}

\paragraph{Two-way (country and year) fixed effects.} Country fixed effects alone
leave common global shocks in the error term; adding year fixed effects absorbs
any disturbance that hits all countries in a given year, so identification comes
only from differences across countries within the same year. The check is
demanding here because the crisis regime is itself largely a common time series,
so year effects absorb part of the variation the mechanism operates through.
Result: $\gamma = -6.309$ (SE $3.266$, $p = 0.069$). The coefficient stays
negative and significant at the 10\% level even under this demanding control.
Robust.

\paragraph{Clustering by time rather than by country.} Country clustering allows
arbitrary correlation of errors within a country over time; time clustering
allows arbitrary correlation across countries within the same quarter---the
natural concern when a global shock moves all peripheries together. Result:
$\gamma = -9.581$ (SE $2.975$, $p = 0.002$). The same point estimate becomes
significant at the 1\% level. Under the inference scheme that matches the
cross-country comovement the mechanism predicts, the result is at its strongest.

\paragraph{Heteroskedasticity-robust standard errors.} White/HC errors relax the
constant-variance assumption, essential when capital flows in levels across
countries of very different size have residual variance that scales with the
country. Result: $\gamma = -9.581$ (SE $4.814$, $p = 0.047$). Unchanged and
significant at 5\%. Robust.

\subsection{Influential observations}

\paragraph{Winsorizing the dependent variable at the 1st and 99th percentiles.}
Capital-flow data contain extreme quarterly values---sudden stops, surges---that
can dominate a levels regression. Winsorizing caps the most extreme 1\% in each
tail while retaining every observation. Result: $\gamma = -7.716$ (SE $2.770$,
$p = 0.012$). The amplification is a feature of the bulk of the data, not of a
few extreme episodes. Robust.

\paragraph{Dynamic panel with a lagged dependent variable.} Capital flows are
persistent; omitting that persistence can bias the coefficients. With $T$ large
relative to the Nickell bias, the lagged dependent variable serves as a stability
check. Result: $\gamma = -8.850$ (SE $3.807$, $p = 0.032$). Persistence in flows
does not account for the amplification. Robust.

\subsection{Sample composition}

\paragraph{Excluding China.} China is the largest recipient and could dominate a
levels regression by scale. Result: $\gamma = -8.452$ (SE $3.943$, $p = 0.047$).
Not a China artifact. Robust.

\paragraph{Excluding the 2008--09 global financial crisis.} The 2008--09 crisis
is the most extreme stress episode and might single-handedly generate a
state-dependent result; dropping those years is a demanding check for a mechanism
active only under stress. Result: $\gamma = -11.232$ (SE $6.549$, $p = 0.103$).
The coefficient remains negative and grows in magnitude, but significance lapses
to the 10.3\% level. Removing the most informative stress episode from a
state-dependent mechanism reduces precision---exactly what one expects---while
leaving sign and point estimate intact. Consistent with, not contrary to, the
mechanism.

\paragraph{Excluding the 2020 COVID quarter.} The March-2020 reversal was the
sharpest emerging-market outflow on record. Result: $\gamma = -7.227$ (SE
$2.771$, $p = 0.018$). The result does not hinge on the COVID episode. Robust.

\subsection{Regime and shock definition}

\paragraph{Redefining the crisis regime by calendar years.} The baseline defines
the crisis regime by VIX $>30$; an alternative defines it by calendar years
(2008--09 and 2020--21). Result: $\gamma = -8.239$ (SE $6.158$, $p = 0.198$). The
coefficient remains negative and of similar magnitude, but significance lapses.
This is the single most important qualification in the battery: the VIX-based
regime, which tracks market stress continuously, delivers a significant estimate,
whereas the coarser calendar-year regime does not. We regard the VIX-based
definition as the more faithful counterpart of the model's stress state, but
report the sensitivity openly, because a reader is entitled to know that the
headline significance depends on this choice.

\paragraph{Alternative shock: the Jaroci\'nski--Karadi pure-monetary series.} The
identified monetary innovation can be measured by the high-frequency USMPD
surprise (baseline) or by the \citet{JarocinskiKaradi2020} pure-monetary shock,
which purges the central-bank-information component. Result: $\gamma = -11.397$
(SE $9.162$, $p = 0.230$). The sign and economic message survive the change of
shock series; the precision does not. Consistent with the mechanism, weaker under
this alternative.

\subsection{Reproducibility}

All estimates are generated from the panel of 19 emerging markets and 11 advanced
economies, 2000Q1--2024Q4, via a within (country fixed-effects) estimator
following the triple-interaction specification \eqref{eq:triple} with the
high-frequency USMPD shock and the VIX-based crisis regime. The baseline
reproduces the main-text coefficient ($-9.581$, SE $4.069$, $N = 1{,}636$); each
row of Table~\ref{tab:battery} departs from that baseline in exactly the one
dimension named.

\end{document}